\documentclass[runningheads]{llncs}
\usepackage[T1]{fontenc}
\usepackage[inline]{enumitem}
\usepackage{booktabs,tabularx}
\usepackage{hyperref}
\usepackage{graphicx}
\usepackage{color}

\begin{document}
\title{Data Product MCP: Chat with your Enterprise Data}
\author{Marco Tonnarelli\inst{1} \and
Filippo Scaramuzza\inst{1} \and
Simon Harrer\inst{2} \and
Linus W. Dietz\inst{3}}
\authorrunning{M. Tonnarelli et al.}
\institute{Eindhoven University of Technology, Tilburg University, 's-Hertogenbosch, Netherlands\\
\email{\{m.tonnarelli,f.scaramuzza\}@tue.nl} \and
Entropy Data GmbH, Monheim am Rhein, Germany\\
\email{simon.harrer@entropy-data.com} \and
King's College London, London, United Kingdom\\
\email{linus.dietz@kcl.ac.uk}}

\maketitle

\begin{abstract}
Computational data governance aims to make the enforcement of governance policies and legal obligations more efficient and reliable. Recent advances in natural language processing and agentic AI offer ways to improve how organizations share and use data.
But many barriers remain. Today's tools require technical skills and multiple roles to discover, request, and query data. Automating data access using enterprise AI agents is limited by the means to discover and autonomously access distributed data. Current solutions either compromise governance or break agentic workflows through manual approvals.
To close this gap, we introduce Data Product MCP integrated in a data product marketplace. This data marketplace, already in use at large enterprises, enables AI agents to find, request, and query enterprise data products while enforcing data contracts in real time without lowering governance standards. The system is built on the Model Context Protocol (MCP) and links the AI-driven marketplace with cloud platforms such as Snowflake, Databricks, and Google Cloud Platform. It supports semantic discovery of data products based on business context, automates access control by validating generated queries against approved business purposes using AI-driven checks, and enforces contracts in real time by blocking unauthorized queries before they run. We assessed the system with feedback from $n=16$ experts in data governance. Our qualitative evaluation demonstrates effectiveness through enterprise scenarios such as customer analytics. The findings suggest that Data Product MCP reduces the technical burden for data analysis without weakening governance, filling a key gap in enterprise AI adoption.
\end{abstract}

\section{Introduction}
Computational data governance lets organizations automate governance by embedding policies and enforcement in code and tools. When done well, it supports value creation and becomes a key part of data management~\cite{wu2024,Mikalef2019BigDA}. However, the many legal obligations, and the gaps in skills, knowledge, and experience between business, technical, and legal teams, make this process more complex~\cite{fadler2020}.
One promising approach to tackle data governance is Data Mesh, a popular socio-technical decentralized data architecture concept based on four principles~\cite{Dehgani2019}:
\begin{enumerate*}[label=\itshape\roman*\upshape)]
    \item \emph{Domain-oriented decentralized data ownership:} Organizes data around specific domains, such as marketing or payments, and makes the teams in those domains responsible for producing and managing their own data.
    \item \emph{Data as a product:} Treats each dataset as a product with clear ownership, service levels, documentation, discoverability, and quality guarantees for its users.
    \item \emph{Self-serve data platform:} Provides a shared platform that lets domain teams publish, find, access, and monitor data without relying on a central team.
    \item \emph{Federated computational governance:} Applies shared standards for security, privacy, and interoperability through a federated model, rules are set centrally but applied and automated within each domain.
\end{enumerate*}
Central platforms such as data catalogs for indexing and retrieving data offerings require large effort to realize~\cite{Labadie2020fair} and data marketplaces are a relatively new concept and thus, still difficult to use for non-technical users. The main reasons are the lack of common definitions, fragmented and siloed data, and immature metadata practices~\cite{wu2024}.
Current products cannot overcome the main barriers, which are the need for technical skills and coordination among multiple roles to discover, request, and query data~\cite{ghosh2025developing}.
As a result, discovering and accessing data products is often slow and time-consuming~\cite{olesen2023enterprise}.
Attempts towards the automation of data discovery and the related governance steps are limited by the ability to locate and retrieve distributed data autonomously. Existing methods either weaken governance or disrupt workflows by adding manual approval steps~\cite{MIKALEF2021103434,mikalef2021artificial}.

The central innovation of our proposed solution is the integration of LLM-powered agents with machine-executable governance. The system uses a natural language chat interface to discover, request, and query data managed in a data marketplace, while maintaining audit trails and moderating data access. In other words, it allows the user to \emph{chat with their enterprise data} in a fully governed manner.
After identifying datasets that fit the analysis, the system checks the declared purpose against the data contracts attached to the products before any computation occurs. Each data access request is logged with the individual data provider along with the declared purpose.
The system generates queries, such as SQL, that respect contracts: schemas, semantic definitions, and rules for joins and aggregations are treated as hard constraints rather than after-the-fact checks. Contract clauses are compiled into executable guards and enforced through a SQL gateway that verifies queries before execution. Queries that violate policies are rejected, and compliant queries proceed.
Each interaction creates an immutable audit chain linking the business question, purpose, governing policies, generated query, and results. This approach moves governance from static documentation and manual review to an executable system that guides both query generation and execution. By building the model's context around high-quality data products and treating contracts as computational objects, the system improves semantic accuracy, reduces policy risks, and ensures compliance without slowing analysts.
\begin{figure}[h]
    \centering
    \includegraphics[width=0.9\linewidth]{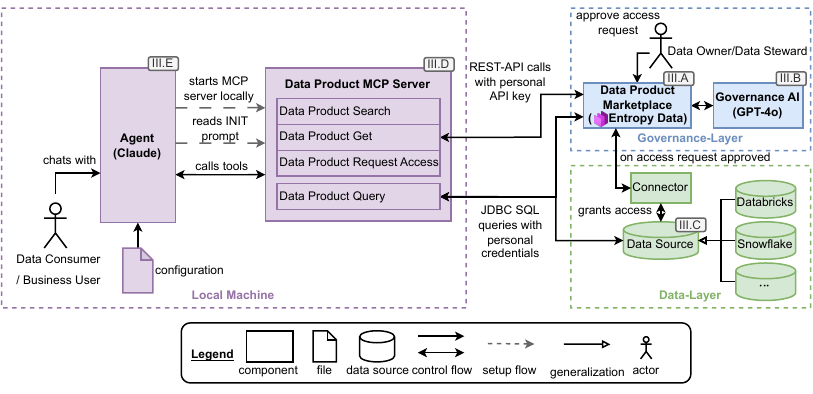}
    \caption{\textbf{System architecture for AI-driven data product access and governance.} The user formulates a business query, for which the Data Product MCP server identifies suitable data sources. Data access is then requested specifically for this query, following governance policies that determine whether approval is automatic or requires review by data owners. Once approved, the MCP server grants controlled access, formulates, and executes the corresponding data queries. The Data Product Marketplace functions as a centralized exchange for data products, ensuring compliance, traceability, and user accountability throughout the entire data lifecycle.}
    \label{fig:architecture}
\end{figure}

This study presents an overview of the proposed solution, shown at a high level in~\autoref{fig:architecture}: an end-to-end agent-based system with a chat interface, built on MCP and a data product marketplace for governance. We evaluated the tool from the perspectives of practitioners and researchers using a qualitative survey based on a demonstration. Participants highlighted its innovative features, particularly for automated AI-enabled governance workflows, supporting non-technical users, and simplifying data product discovery.

\section{Background and Related Work}
\subsection{Data Mesh}
\begin{figure}
    \centering
    \includegraphics[width=1\linewidth]{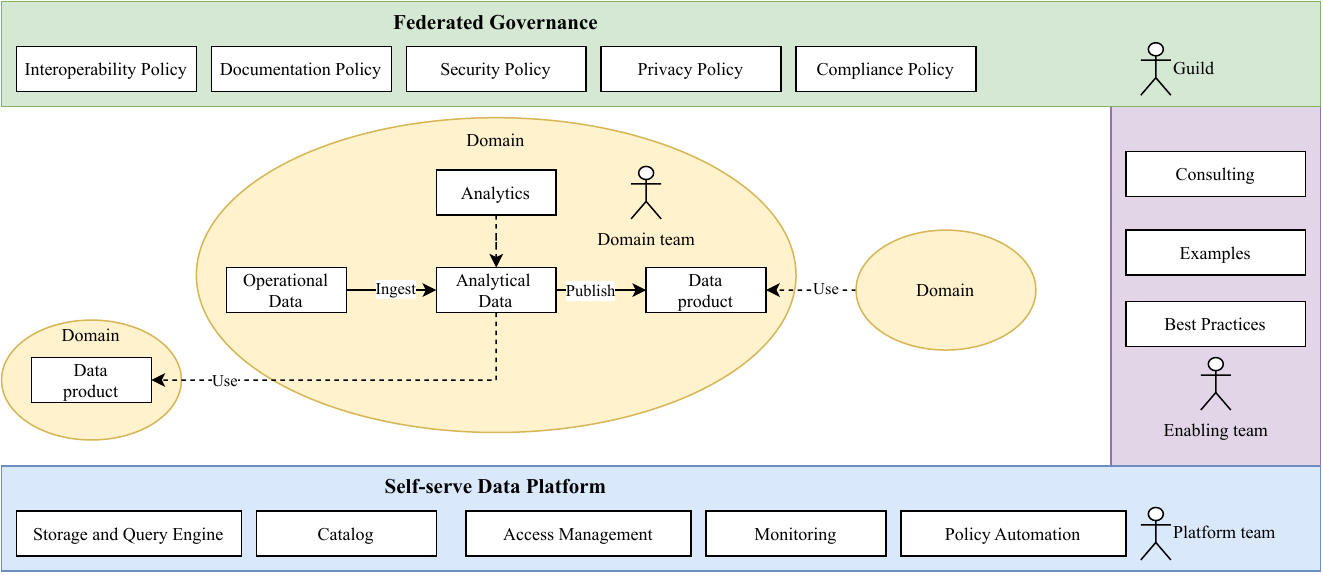}
    \caption{Data Mesh and its principles.}
    \label{fig:datamesh}
\end{figure}

To address the architectural and organizational bottlenecks of centralized data management architectures, the industry has increasingly adopted Data Mesh~\cite{dehghani2020data}, a decentralized, socio-technical architecture driven by four foundational principles: (i) Domain-oriented decentralized data ownership, (ii) Data as a product, (iii) Self-serve data platform, (iv) Federated computational governance. Fig.~\ref{fig:datamesh} shows an overview of data mesh and its principles. 
With domain-oriented decentralized data ownership, accountability and data management are distributed to specific business domains (e.g., marketing, payments) rather than a central data team; Data as a product treats datasets as formalized products requiring clear ownership, guaranteed service-level agreements (SLAs), comprehensive documentation, and strict quality guarantees. The self-serve data platform is a shared, underlying infrastructure that enables autonomous domain teams to publish, discover, and consume data without centralized engineering dependencies. Finally, federated computational governance centrally defines global standards for security, privacy, and interoperability, but it enforces them automatically within individual domains.

\subsection{Data Catalogs and Data Marketplaces}
The success of a Data Mesh relies heavily on robust metadata management to ensure discoverability and interoperability. Historically, this has been handled by Data Catalogs, which act as centralized indexes of an organization's data assets~\cite{TONNARELLI2025112584}. However, realizing a functional data catalog requires massive organizational effort, and they often remain inaccessible to non-technical users due to complex user interfaces, siloed infrastructure, and immature metadata standards~\cite{jahnke2023}. Data Marketplaces represent the next evolution, shifting the focus from mere asset indexing to active exchange. A marketplace facilitates the sharing of data products between producers and consumers while inherently enforcing governance functions, such as data quality controls, global policies, and formal data contracts~\cite{driessen2022}. Despite these advancements, current marketplace solutions still struggle with friction; discovering, requesting, and querying data products frequently demands technical proficiency and slow, manual coordination across multiple roles~\cite{abbas2021business}.

\subsection{AI-Assisted Governance}
Recent advancements in natural language processing and agentic AI offer pathways to streamline data sharing~\cite{jabal2020flapfederatedlearning}. Researchers have proposed LLM-based assistants capable of reviewing user data access requests and evaluating them against privacy policies and formal data contracts. However, fully automating data access via enterprise AI agents faces a critical barrier: the inability to autonomously locate and retrieve distributed data without compromising security~\cite{elluri2021}. Existing automation attempts either bypass strict governance protocols or disrupt the agent's workflow by forcing manual, human-in-the-loop approval steps~\cite{cejas2023}. True computational governance requires that data contracts (schemas, semantic definitions, usage limitations) are treated as machine-executable hard constraints, blocking unauthorized queries at the gateway before any computation occurs~\cite{JUSSEN2024102280}~\cite{waithira2019data}.
\subsection{The Model Context Protocol}
The integration of LLMs with internal enterprise infrastructure has been significantly accelerated by the MCP~\cite{mcp_intro}. MCP is an open, standardized protocol that allows large language models to securely interact with external data sources, tools, and services. While the industry has rapidly adopted MCP for use cases in software engineering and infrastructure management, academic research has yet to explore its integration deeply. Specifically, no existing implementations have applied MCP to solve the complex problem of semantic data product search, automated access requests, and governed query execution within a decentralized enterprise data marketplace.

\subsection{Related Work}

Literature on data mesh focuses on its principles, architectures and governance challenges. A substantial body of work examines the four foundational principles of data mesh and highlights how they enable modularity, scalability, and effective governance across distributed data ecosystems governed by these tenets~\cite{machado2021data,machado2022arch,vander2024,sona2024,blohm2024}. To support adoption, researchers have introduced multiple reference architectures that present alternative archetypes for implementing data mesh within organizations~\cite{vander2024,schleicher2024data,Goedegebuure2024,borovits2024}.

Metadata management enables discoverability, interoperability, and governance in data product–oriented ecosystems~\cite{Goedegebuure2024}. Prior work proposes formal metadata models to manage data products across domains~\cite{Driessen2023Metadata,Driessen2023ProMoTe,wider2023}. To improve automation in governance and policy enforcement, Dietz et al.~\cite{Dietz2025Automating} present an LLM-based assistant that reviews data access requests in a data mesh environment against privacy policies and data contracts.
Self-serve platforms are central to data mesh adoption, with research describing infrastructure elements such as APIs, data catalogs, and processing engines~\cite{vaneijk2024,sona2024}. However, tools like data catalogs remain difficult for most non-technical users to use because of their complexity, and researchers advocate simpler ways to interact with them~\cite{olesen2023enterprise}.

A core capability of Data Product MCP is translating natural language questions into SQL queries, placing it in the broader landscape of Text-to-SQL research. Recent work has shown that LLMs struggle with real-world enterprise SQL tasks: Spider~2.0~\cite{lei2025spider2} benchmarks state-of-the-art models on enterprise workflows across BigQuery and Snowflake and finds that only 21.3\% of tasks are solved correctly. Surveys of NL2SQL systems~\cite{shi2025surveyllmtexttosql} and benchmarks on knowledge-graph-augmented enterprise SQL question answering~\cite{sequeda2023benchmark} confirm that schema complexity and domain-specific semantics remain key challenges. Data Product MCP addresses these challenges by grounding SQL generation in machine-readable data contracts that supply precise schema definitions, join rules, and semantic annotations as hard constraints.

Within this context, the Model Context Protocol has been applied to address similar challenges. MCP is a standardized, open protocol that allows LLMs to interact with external data sources and services~\cite{mcp_intro}. The concept has quickly gained interest in industry, with documented applications across several domains, including development and engineering~\cite{noauthor_githubgithub-mcp-server_2025}, infrastructure management~\cite{noauthor_cloudflaremcp-server-cloudflare_2025,noauthor_gomarble-aifacebook-ads-mcp-server_2025}, and data management and analytics~\cite{noauthor_snowflake-labsmcp_2025,atlanhq2025mcp,googleapis2025toolbox,dbtlabs2025mcp}, among others.
The Atlan MCP Server~\cite{atlanhq2025mcp} exposes a metadata catalog with tools for asset search, lineage traversal, and data quality rules, but provides no data contract enforcement or purpose-based access control. Google's MCP Toolbox for Databases~\cite{googleapis2025toolbox} offers broad connectivity to cloud databases (including BigQuery and Snowflake) for SQL execution, yet it lacks any marketplace, governance layer, or access-request workflow. The dbt MCP Server~\cite{dbtlabs2025mcp} integrates with dbt semantic layers to support natural-language metric queries, but is scoped to the dbt transformation stack and does not address data product discovery or governance.
Academic research on MCP is now emerging. An empirical study of 177,000 MCP tools~\cite{stein2026mcp} finds that software development accounts for 67\% of current tools, while action-oriented tools are rising rapidly. Mansouri et al.~\cite{mansouri2026mcp} propose an MCP server that exposes financial data APIs for question answering, demonstrating 80.4\% accuracy and positioning MCP as a viable alternative to RAG for structured data, an approach directly analogous to ours. Dobriy et al.~\cite{dobriy2026sparql} present an MCP-based SPARQL agent for federated knowledge graph querying, showing the protocol's applicability to distributed data access.
Of particular relevance to our work, no existing MCP implementation combines data product discovery, purpose-based access control, data contract enforcement, and governed SQL query execution within a single endpoint.

\section{System Architecture}
The system allowing the user to \textit{chat with their enterprise data} is built on two pillars: data products and LLMs. Data products provide semantic context through metadata and data contracts. This information allows the AI to understand what data exist, what they mean, and how to use them correctly, according to the terms of the data contracts and global governance policies.
To realize this, we designed a system architecture (\autoref{fig:architecture}) using five core components: the \emph{Data Product Marketplace} (\autoref{subsec:dpm}), the \emph{Governance AI} (\autoref{subsec:aigov}), the \emph{data sources} (\autoref{subsec:data_sources}), the \emph{Data Product MCP Server} (\autoref{subsec:DPMCP}), and the \emph{LLM-based Agent Chat Interface} (\autoref{subsec:chatbot}). %

The Data Product Marketplace serves as both a catalog and a marketplace for discovering data products and enforcing governance. It supports interfaces with well-established enterprise data sources, including Databricks. The data catalog is a centralized inventory of an organization's data assets, while a data marketplace adds features for sharing data between producers and consumers along with governance functions such as policies, standards, and data quality controls.
To automate data governance, LLMs handle data product access requests by using data contracts and the declared purpose of access to enforce policies automatically. Users submit these requests through a chat interface, such as Anthropic Claude, which interacts with our MCP server. %
The natural language requests are translated into SQL queries, annotated with their purpose, and sent to the Data Product Marketplace, where governance is enforced, and finally to the data sources, where the data is stored. The next sections describe these five components in detail.

\subsection{Data Product Marketplace}
\label{subsec:dpm}
We use Entropy Data\footnote{\url{https://www.entropy-data.com}}, formerly known as Data Mesh Manager, as the data product marketplace to manage data products, data contracts, and global governance policies~\cite{Wider2025AI}.
Data product owners can define their products and contracts in the studio using open standards such as the Open Data Product Standard (ODPS)~\cite{ODPS2025} and the Open Data Contract Standard (ODCS)~\cite{ODCS2025}, and then publish them in the marketplace for data consumers to discover and request access. Owners can either approve access requests manually or specify automatic approval rules, for example, for requests within teams or for non-critical data.
The data product marketplace stores only metadata; the actual data remain in the connected data sources. As a result, queries are executed at the data source providers.

\textbf{Core Innovation:} The platform centrally tracks who has access to which data, for what purpose, and under what terms and conditions defined in data contracts and global policies.

\subsection{Governance AI}
\label{subsec:aigov}
Governance AI is an LLM-based assistive control system designed to automatically ensure policy compliance~\cite{Dietz2025Automating}. When access to a data product is requested, the system analyzes the relevant data contracts and global policies to determine whether the request is permissible. Based on the declared purpose of access and the applicable policies, it produces structured warnings (e.g., potential violations of PII protection) to assist data product owners in maintaining compliance.

Technically, Governance AI uses a GPT model through a platform API. A system prompt enforces a structured review process, while a user prompt provides all necessary artifacts, such as products, contracts, teams, and policies in a structured format. The resulting analysis is presented to the data product owner within the Data Product Marketplace, where access approvals or rejections are ultimately recorded.
Under the EU AI Act, this system is classified as \emph{limited risk}, requiring that users be informed that the warnings are generated by an AI system.

\textbf{Core Innovation:} The system automatically checks whether a given purpose complies with the terms and conditions and policies written in natural language.

\subsection{Data Sources}
\label{subsec:data_sources}
The system, in particular the data product marketplace and the Data Product MCP server, integrates with several data sources and platforms such as Databricks, Snowflake, Google Cloud Platform, Microsoft Azure, and Amazon Web Services.
Integration with these services is implemented through dedicated connectors that provide APIs for interaction. These platforms store the data for registered data products and are accessed by the Data Product MCP Server when a query is submitted. The two main features of these integrations are: \emph{(i)} Synchronizing data assets, including tables and schemas, from the source to the data product marketplace; \emph{(ii)} managing data access for data consumers.

\subsection{Data Product MCP Server}
\label{subsec:DPMCP}
The Data Product MCP~\cite{noauthor_entropydata-dataproductmcp-server_2025} is an open source\footnote{Source code available at \url{https://github.com/entropy-data/dataproduct-mcp}} MCP server written in Python using FastMCP that enables the discovery of data products, access requests in Entropy Data, the data product marketplace, and query execution on data platforms that host the actual business data. It receives requests from an agent and processes them using four tools: \texttt{Data Product Search}, \texttt{Data Product Get}, \texttt{Data Product Request Access}, and \texttt{Data Product Query}.

\subsubsection{\texttt{Data Product Search}}
This tool offers a query-based search for data products, returning only a summary of a data product. Internally, the tool uses a single REST API call with filtering options, such as returning only active products, to limit the retrieved information. It supports multiple search methods, including list-based and semantic search. Search terms can be applied to the product ID, title, and description, with support for multiple terms. The output is a structured list containing each product's ID, name, description, and owner.

\subsubsection{\texttt{Data Product Get}}
This tool retrieves the full details of a data product by its ID. Through API calls to the data product marketplace, it collects details about the data product, connection details, access status information, and associated data contracts, all merged into a single response to the agent.

\subsubsection{\texttt{Data Product Request Access}}
This tool requests access to a specific data product. It does this via a single API call to the data product marketplace. The request may be granted automatically or require manual review by the data product owner, depending on the product's configuration, declared purpose, and governance rules.

The required inputs are the data product ID and purpose of use. If review by the data provider is required, the declared purpose helps the data owner assess eligibility from business, technical, and governance perspectives. The tool returns the access request details, including its status and approval information, and logs the access request for audit reasons.

\subsubsection{\texttt{Data Product Query}}
This tool performs several API calls before formulating and executing a SQL query. It first contacts the data product marketplace to retrieve metadata about the data product including connection details. It then verifies whether the user has active access to the data product and, if true, runs an access evaluation via Governance AI to ensure the query complies with governance rules. After these checks, the tool executes a single query on the underlying data platform, returning the results back to the agent.

\textbf{Core Innovation:} The combination of data marketplace, governance and query capabilities as a single MCP endpoint to allow querying suitable data products within the organization using natural language.

\subsection{Agent}
\label{subsec:chatbot}
The chat interface of the Agent is the primary point of interaction between the end-user and the system. Within the current implementation, the agent employed is Claude Desktop, by Anthropic. However, the architecture is not limited to this specific model; in principle, any chat-based LLM supporting MCP could be integrated with the system.

The agent requires a configuration file as input. This file specifies the commands needed to initialize and configure the Data Product MCP server, as well as environment variables. These variables include private API keys that uniquely identify a user and define their role-based privileges, which are required to access the data product marketplace and connected data sources.

The agent processes natural language input from the end-user and determines whether the Data Product MCP server should be invoked. When the server is engaged, the message is relayed accordingly, and all subsequent actions performed by the MCP server are presented transparently within the chat interface, along with sufficient contextual information for interpretability. Depending on the nature of the query, the system delivers results to the end-user in a structured and accessible manner.

\textbf{Core Innovation:} The user provides the Agentic AI with a business question, which it then addresses using the tools available through the Data Product MCP.

\section{Scenario-based Demonstration of Data Product MCP}\label{sec:scenarios} %
This section illustrates the functionality of the system through three representative scenarios: identifying top customers with the human in the loop (\autoref{uc:top-customers}), analyzing support cases without additional human oversight (\autoref{uc:support_cases}), and preventing misuse of customer data at query time (\autoref{uc:miuse}).

\begin{figure}[p]
    \centering
    \includegraphics[width=\linewidth]{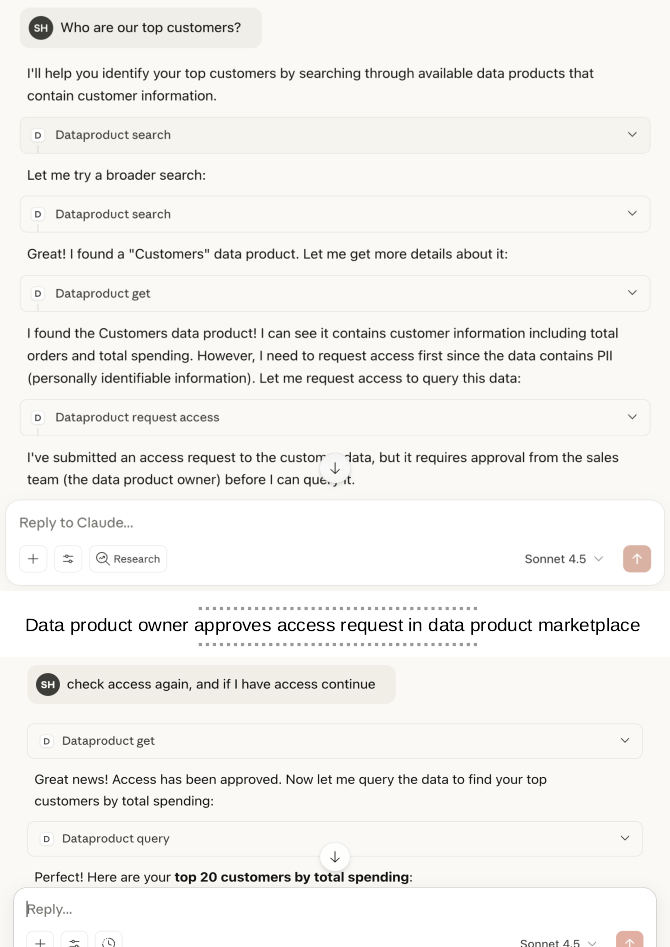}
    \caption{Identifying Top Customers with the Human in the Loop} %
    \label{fig:uc1}
\end{figure}

\subsection{Identifying Top Customers with the Human in the Loop}
\label{uc:top-customers}
To illustrate the system's functionality, we consider the question: \emph{``Who are our top customers?''} as shown in~\autoref{fig:uc1}. The agent begins with a semantic search for relevant data products. An initial narrow query (``customer sales revenue'') returns no direct results, so the agent broadens the search (``customer'') and discovers the customer data product. Based on a brief summary, the agent retrieves all metadata for the product, including the associated data contract and whether the current user already has access. At this stage, the user does not yet have access to the data product.

The agent derives the intended purpose of use from the user's initial query and
then initiates a formal access request in the Data Product Marketplace via the Data Product MCP. Because in this case, a manual approval workflow is in place, the agent cannot proceed further. The responsible data product owner is notified by the marketplace and, as the human in the loop, will be asked to assess the request through the web interface.

After approval, the user asks the agent to check access and proceed. The agent retrieves the data product metadata again, confirms that the access request has been approved, and continues with the query to identify the top customers. In the background, the Data Product MCP verifies that the query's purpose aligns with governance policies, the data contract's terms and conditions, and the purpose under which access was granted. Once confirmed, the MCP executes the query on the underlying data platform, in this case, Databricks, and returns the results to the user, effectively answering the original business question.

\subsection{Analyzing Support Cases Without Additional Human Oversight}
\label{uc:support_cases}

A second scenario considers the question: \emph{``What are the top reasons for support tickets?''} Here, the agent locates a data product containing support ticket records. Unlike the previous example, this data product is classified as non-sensitive and is subject to an automated approval workflow. As a result, the access request is granted immediately without human intervention; still, all actions are logged to ensure auditability.

Once access is established, the agent generates and executes SQL queries against the data product. In this scenario, the queries are more complex, involving advanced operations such as window functions, but they are automatically generated in compliance with the data contract metadata. The process remains fully governed, with each query linked to its declared purpose. Ultimately, the agent provides a structured answer to the user's question.

\subsection{Preventing Misuse of Customer Data at Query Time}
\label{uc:miuse}

The third scenario highlights how the system enforces governance by rejecting queries that are misaligned with approved business purposes. Consider the request: \emph{``Create a CSV file of all our top customers for a luxury email campaign''}. The agent successfully interprets the query intent and associates it with the declared purpose of marketing outreach.

\begin{figure}[h]
    \centering
    \includegraphics[width=\linewidth]{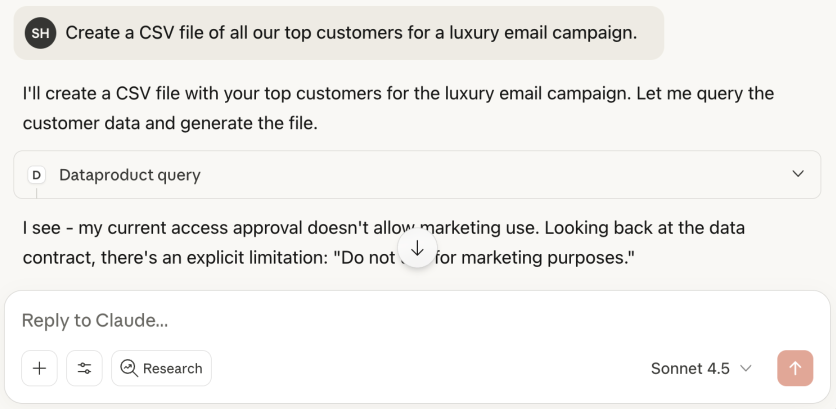}
    \caption{Preventing Misuse of Customer Data at Query Time} %
    \label{fig:uc3}
\end{figure}

However, the Data Product MCP, with the help of Governance AI, determines that the intended use case is not permitted, as shown in~\autoref{fig:uc3}. Although the user has access to the underlying customer data product, the MCP blocks query execution because the stated purpose does not comply with the terms and conditions of the data contract.

This example illustrates the central principle of purpose-based governance: data access alone is insufficient to justify use. Each query must align with the contractual and policy framework that governs the data product. By enforcing this constraint, the system ensures that even valid credentials cannot be misused for unauthorized business activities.

\section{Evaluation}

We evaluated the innovative potential of Data Product MCP using the three scenarios from~\autoref{sec:scenarios}. We chose this approach to get opinions from a wider audience and to ensure that the findings are not limited to individual organizations. Thus, we showed a demonstration to participants at an industry conference and a webinar promoted on LinkedIn. After the demonstration, we invited participants to join a scientific study and shared a survey link for those who wished to take part.
The survey gathered participants' job titles, reflecting the diverse expertise of the 16 respondents. The largest group consisted of product and engineering management roles ($n=5$), followed by data architects ($4$), executives in C-level positions ($2$), directors or heads ($3$), and academic researchers ($2$).
To collate the impressions and concerns of the participants, we used a qualitative approach focused on thematic analysis to summarize the answers~\cite{miles1994qualitative,braun2006using}.

\medskip
\noindent\textbf{After seeing the demo, what tasks or use cases in your work do you think Data Product MCP enables that you could not do (or could do only with much more effort) before?}
Experts identified three main benefits: natural-language access to data, stronger governance, and quick and accessible self-service. Many emphasized the possibility to obtain quick exploratory insights without deep technical skills and knowing which data products exist within the data mesh. As the system can translate business questions into runnable queries, it considerably reduces the entry barrier to consuming data. As one expert put it: \emph{``[It] makes it very data-democratic, as just asking questions is a nice abstraction (E5).''}

The automatic integration of governance was mentioned as a differentiator (E1, 3, 6--7). Experts cited the purpose validation, audit trails tied to purpose, automated access requests, and governance applied to data product and APIs. As one expert put it, these features enable \emph{``Conversational AI in a governed manner (E1)''} and shorten the path from discovery to consumption.

A common interest was expressed on the ability to run cross-data-product queries and compose results with data fusion and subsequent transformations (E4--5, 10, 12).
Other niche points included just-in-time context management (E12) to avoid large-language-model limits and out-of-the-box platform integration.

\medskip
\noindent\textbf{How do you expect the use of Data Product MCP to change your current approach to data work (e.g. adoption of Data Mesh, relevance of dashboards, collaboration across teams)?}

Experts expected the Data Product MCP to speed up and broaden access to insights, shifting work from dashboard-led business intelligence to conversational, search-driven analysis. They predicted a decrease in reliance on traditional tools that could lower licence costs (E1). Others predicted stronger governance, traceability and audit trails that let users explore data openly while keeping control and meeting compliance rules (E7, 9).

Several respondents thought it will advance practices aligned with the Data Mesh, decentralized, domain-led data ownership, by making cross-domain reuse and composition easier and by building a shared vocabulary across teams, which should improve collaboration and discovery (E4, 6, 10--11, 16). Others added it will encourage higher-quality, consumer-friendly data products and shorten the exploration phase, so teams find valuable assets faster (E1--2, 6, 13--14).

A common theme was that agents will be able to tap and act on large data catalogs, making it easier to prototype and productize such use cases and generally foster adoption of data mesh architectures within the corporation (E3--6, 11--12, 16).

\medskip
\noindent\textbf{How reliable would Data Product MCP's results need to be for you to trust and adopt it in production?}
Most experts set a high bar for production use based on the following distribution of answers on a Likert scale from Low (can tolerate frequent errors if it still saves time, $n=0$) -- Moderate (some errors acceptable with human review, $n=2$) -- Very high (small, rare errors acceptable, $n=10$) --  Near 100\% (critical: no room for errors $n=3$); Not sure, $n=1$).
While no respondent backed low reliability, interestingly only two selected near 100\% as condition for adopting the technology. Furthermore, the consensus among all corporate decision makers (C-level and directors) was that small and rare errors are acceptable.

\medskip
\noindent\textbf{What potential limitations or concerns (e.g. scalability, correctness, security, compliance) do you see that might prevent adoption in your organization?}
Experts identified four main adoption risks.
\begin{description}
\item{\textbf{Data and metadata quality.}} Good input is the basis of correct, trustworthy output (E1). Incomplete metadata and complex prompts can undermine results, and users will need training to spot when outputs are unreliable (E7).

\item{\textbf{Security, identity and access controls.}} Respondents stressed reliable access management, avoiding letting large language models make security decisions, separating human and AI identities, the ability to limit AI permissions, and strong audit logs and traceability (E6, 10--11, 13).

\item{\textbf{Scalability, performance and cost control.}} Concerns included server bottlenecks, handling large metadata volumes, worse performance than direct queries to data platforms, and cost spikes from heavy query loads (E2, 7--9, 14).

\item{\textbf{Compliance and trust.}}
Respondents mentioned auditable results (E3, 5), trust in an LLM-based system (E5, 10, 14) due to the probabilistic nature of the technology and potential hallucinations.
\end{description}
Further aspects mentioned were organizational resistance to forcing a single access channel (E8) and demands to restrict use to local models or tightly controlled interfaces (E14). Finally, one expert was unsure about the automatic query generation aspect: \textit{``I would be very sceptical in it finding the exact right columns for my usecase, especially if the usecase is less standard. (E15)''}

\section{Discussion}

Our initial evaluation of Data Product MCP shows both practical potential and theoretical importance in changing how organizations use data. Experts across roles identified three key capabilities: natural-language access, faster self-service and embedded governance.
In practice, the system lowers the barrier to analytics by turning business questions into executable queries. This ``data-democratic'' approach reduces reliance on specialized skills and legacy dashboards, which can shorten decision cycles and lower software costs. Yet, simplified access to an organization's data landscape comes with governance challenges. The system's embedded governance features address this tension by embedding compliance into the process. Governance thus moves from a regulatory requirement users need to adhere to an enabling technical layer that supports transparent and auditable exploration of data.

Trust and reliability emerged as critical for adoption. Most respondents required high but not perfect accuracy, suggesting that transparency and traceability of errors are more important than absolute precision.
Concerns regarding data quality, scalability, security, and compliance reflect broader organizational challenges with data. They also highlight a tension between automation and control: while AI systems mediate user intent and query execution, accountability must remain under human oversight.

Our approach aims to shift business data from a specialized technical asset to a widely available resource. For this reason, its success and adoption depend not only on technical maturity but also on cultivating a culture of high-quality data sharing across the organization. If this is achieved, Data Product MCP can provide fast, effective, and governed access to data ecosystems and enhance organizational intelligence.

\section{Conclusions}

We introduced the Data Product MCP, which enables chat-based access to enterprise data through data product discovery while enforcing data governance. By leveraging LLM-powered agents and the capabilities of a data product marketplace within federated data architectures, the system removes barriers such as technical expertise requirements and manual oversight. Our qualitative evaluation, based on scenario-driven demonstrations, showed that it supports both human-in-the-loop and fully automated queries while maintaining governance compliance.
Feedback from 16 experts highlighted key benefits, including natural language access, faster exploratory analysis, stronger governance, and improved cross-domain collaboration, as envisioned by Data Mesh principles. Remaining challenges include maintaining metadata quality, ensuring security, improving scalability, and fostering trust in AI-generated outputs.

\bibliographystyle{splncs04}
\bibliography{literature.bib}

\end{document}